\documentclass[showpacs,twocolumn,prl]{revtex4}
\usepackage{epsfig}
\usepackage{amsmath}
\usepackage{graphicx}
\usepackage{amssymb} 
\usepackage{graphics}

\usepackage{fancyhdr,graphicx}

\begin{document}

\title{Silicon-based nanochannel glucose sensor}

\author{Xihua Wang$^1$, Yu Chen$^1$, Katherine A. Gibney$^2$, Shyamsunder Erramilli$^1$ and Pritiraj Mohanty$^1$}

\affiliation{
$^1$ Department of Physics, Boston University, 590 Commonwealth Avenue, Boston, MA 02215\\
$^2$ Department of Chemistry, University of Michigan, 930 N. University, Ann Arbor, MI 48109-1055}


\begin{abstract}
Silicon nanochannel biological field effect transistors have been developed for glucose detection. The device is nanofabricated from a silicon-on-insulator wafer with a top-down approach and surface functionalized with glucose oxidase. The differential conductance of silicon nanowires, tuned with source-drain bias voltage, is demonstrated to be sensitive to the biocatalyzed oxidation of glucose. The glucose biosensor response is linear in the 0.5-8 mM concentration range with 3-5 min response time. This silicon nanochannel-based glucose biosensor technology offers the possibility of high density, high quality glucose biosensor integration with silicon-based circuitry.
   
\end{abstract}


\maketitle

\thispagestyle{fancy}
\renewcommand{\headrulewidth}{0pt}
\fancyhead[c]{Appl. Phys. Lett. 92, 013903 (2008)}    

Field effect devices, such as capacitive electrolyteinsulator-semiconductor sensor, light-addressable potentiometric sensor, and ion-sensitive field effect transistor (ISFET) for glucose detection\cite{Karube98,Schoning06,Soldatkin99}, have been extensively studied in recent years. Although these devices are limited by the dependence of the sensor response on buffer capacity, ionic strength, and pH of the test sample, their compatibility with
advanced microfabrication technology may enable their potential commercialization. Glucose biosensor is a particularly attractive enzyme biosensor due to its potential widespread use in clinical applications.
Currently, glucose detection is mostly limited to in vitro test of blood samples, although it is more meaningful to perform in vivo test by implantable sensing devices for continued monitoring of blood glucose level. To this end, nanoscale sensors may be fundamentally valuable. Semiconductor nanowires, grown from bottom-up approach, have been demonstrated as good candidates for ultrasensitive biosensors in many applications\cite{Lieber01,Lieber05, Dekker03}. However, most of the existing studies based on bottom-up approaches face the limitation of complex integration and scalable large-scale manufacturing. Fabrication of silicon nanoscale devices with top-down approaches\cite{Mohanty06,Reed07} using lithography has the inherent benefit of
standard semiconductor processes and, hence, better control of device properties. Therefore, field effect devices, using nanoscale technology, offer the possibility of highperformance, low-cost implantable glucose biosensors.

\begin{figure} [t]
	\includegraphics[scale=0.4]{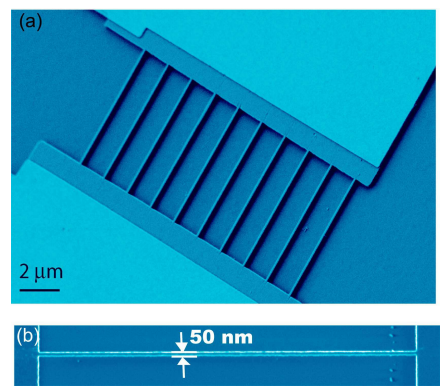}
	
	\caption{ (a) Scanning electron micrograph of silicon nanowires, the working part of the biosensor. (b) A 50 nm wide nanowire image, seeing from top view, shows the controlled linear geometry.  }
	
	\label{fig1}
\end{figure}
Here, we demonstrate silicon nanochannels, surface functionalized with glucose oxidase, as a field effect glucose biosensor. The glucose biosensor with a set of silicon nanowires as nanochannels, 50-100 nm wide, 100 nm high, an6 $\mu$m long, is fabricated from a silicon-on-insulator (SOI) wafer. Figure ~\ref{fig1} shows a typical scanning electron micrograph of the device. The SOI wafer consists of a 100 nm thick silicon top layer, a 500 $\mu$m silicon substrate, and a 380 nm SiO$_2$ insulation layer in between. The silicon top layer is lightly doped with boron concentration, $10^{-15}$ cm$^{-3}$, as the device layer. The silicon nanowires are patterned with electron beam lithography, and two sidewalls are exposed after reactive ion etching. The Ti/Au layer is deposited with thermal evaporator to function as the source and drain contact electrodes without further annealing. The silicon nanowires are covered with a layer of 10 nm Al$_2$O$_3$, grown by atomic layer deposition, to prevent current leakage between analyte solution and silicon nanowires.

Before modification, the Al$_2$O$_3$ surface is treated with oxygen plasma \cite{Williams04} for two purposes: To clean the sample surfaces and to generate a hydrophilic surface. The wires are first modified by APTES (3\% in ethanol with 5\% water). Then 3\% glucose oxidase in acetic chloride (50 mM) buffer (pH 5.1) (5\% glycerol, 5\% BSA) is deposited on the sample and kept in glutaraldehyde vapor for 40 min. The sample is dried in air for 15 min. Glucose samples are made in solution with 50 mM NaCl and 50 mM (or 81.8 mM) of potassium ferricyanide. 

All electrical measurements are done at room temperature. The voltage V across the nanowires is applied by using a modulated sine wave source upon dc bias. The current I accross nanowires is converted into voltage by a feedback resistor, and this voltage is detected with a lock-in amplifier. A voltage is applied on the Ag/AgCl reference electrode contacted with solution as the reference gate voltage, V$_{rg}$ \cite{Mohanty06}. This allows direct, real-time measurement of the differential conductance $g=\left(\frac{\partial I}{\partial V}\right)_{V_{rg}}$. In particular, the method allows studies at zero or reverse bias as well.

\begin{figure} [t]
	\includegraphics[scale=0.75]{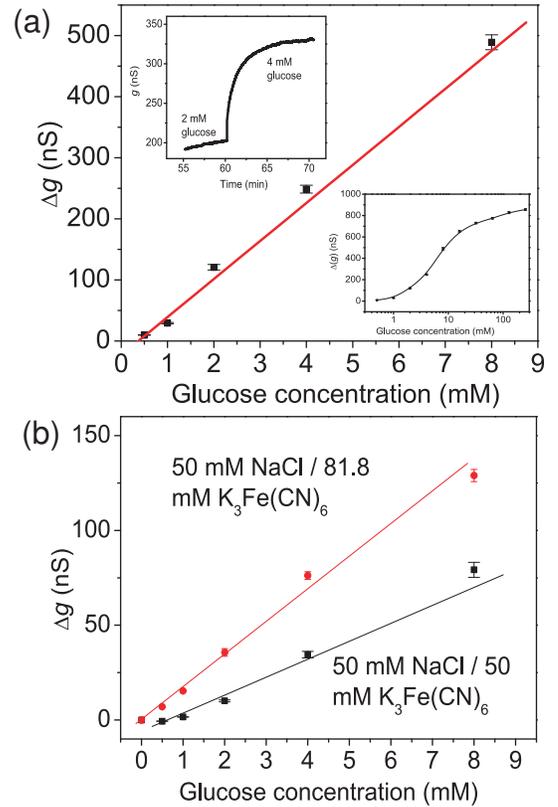}
	
	\caption{(a) The glucose biosensor response in the 0.5 mM - 8 mM concentration range, which is fitted into a linear curve. $V_{ds}=-0.69$ V, $V_{rg}=0$ V. All solution used contents 50 mM of NaCl and 50 mM of potassium ferricyanide. The top inset shows the real time differential conductance change due to exchange glucose solutions.  The bottom inset is differential conductance of the device vs. glucose concentration plot. The random error is estimated to be 5\% of device response due to adding different glucose solutions. (b) Device responses for 81.8 mM (red curve) and 50 mM (black curve) potassium ferricyanide solutions. 	}
	
	\label{fig2}
\end{figure}

Fig.~\ref{fig2} (a) gives the real-time device response when we add glucose solutions upon initial use. We find that there is a linear relation between device response and glucose concentration in the range of 0.5-8 mM. From the response curve (the bottom inset of Fig.~\ref{fig2} (a)), we find that the differential conductance change of the device reaches saturation after the glucose concentration of 16 mM. We can also estimate the diffusion-limited  time response of the device to be 3-5 min, which is the same as the typical response times of thick-layer glucose oxidase modified ISFET \cite{Kwon97}. The response time could be further improved by enzyme monolayer functionalization \cite{Willner00}.  
The oxidation of glucose catalyzed by glucose oxidase is a two-substrate reaction \cite{Rooij94}. One molecule of glucose oxidase contains two molecules of oxidized form of flavine adenine dinucleotide (FAD). In the first substrate reaction, there is a formation of an enzyme-glucose complex, and then reduce the FAD according to
\begin{equation}
	\mbox{C}_6\mbox{H}_{12}\mbox{O}_6(\mbox{glucose})+\mbox{FAD}=\mbox{C}_6\mbox{H}_{10}\mbox{O}_6+\mbox{FADH}_2.
\end{equation}

\fancyhead[c]{}    
\pagestyle{fancy}
\rhead[R]{X. Wang et al, Appl. Phys. Lett. 92, 013903 (2008)}

\noindent The reduced form of the enzyme is fast oxidized by oxygen, producing hydrogen peroxide, and it restores the initial state of the enzyme molecule
\begin{equation}
	\mbox{FADH}_2+\mbox{O}_2=\mbox{FAD}+\mbox{H}_2\mbox{O}_2.
\end{equation}

\noindent In the second substrate reaction, the generated gluconolactone (C$_6$H$_{10}$O$_6$) is hydrolyzed spontaneously to gluconic acid
\begin{equation}
\mbox{C}_6\mbox{H}_{10}\mbox{O}_6+\mbox{H}_2\mbox{O}_2=\mbox{C}_6\mbox{H}_{11}\mbox{O}_7(\mbox{gluconate})+\mbox{H}^+.
\end{equation}

\noindent Therefor, one glucose molecule yields one hydrogen ion, and it changes the local hydrogen ion activity of the solution. In our earlier work, we have shown that the top-down fabricated sensors are sensitive to pH, which quantifies the hydrogen ion concentration. Changes in the local hydrogen ion concentration alter the surface potential, and hence the electric field which modualtes the conductance of the BioFET. This field effect is further amplified by adding ferricyanide ions ([Fe(CN)$_6$]$^{3-}$):
\begin{equation}
	\mbox{FADH}_2+\mbox{2[Fe(CN)$_6$]$^{3-}$}=\mbox{FAD}+\mbox{2[Fe(CN)$_6$]$^{4-}$}+2\mbox{H}^+.
\end{equation}

\noindent Fig.~\ref{fig2} (b) confirms this effect. An 81.8 mM potassium ferricyanide solution has the sensitivity of 16 nS/mM, while a 50 mM potassium ferricyanide solution gives the sensitivity of 10 nS/mM. So higher concentration of potassium ferricyanide yields higher sensitivity. 

\begin{figure} [t]
	\includegraphics[scale=0.5]{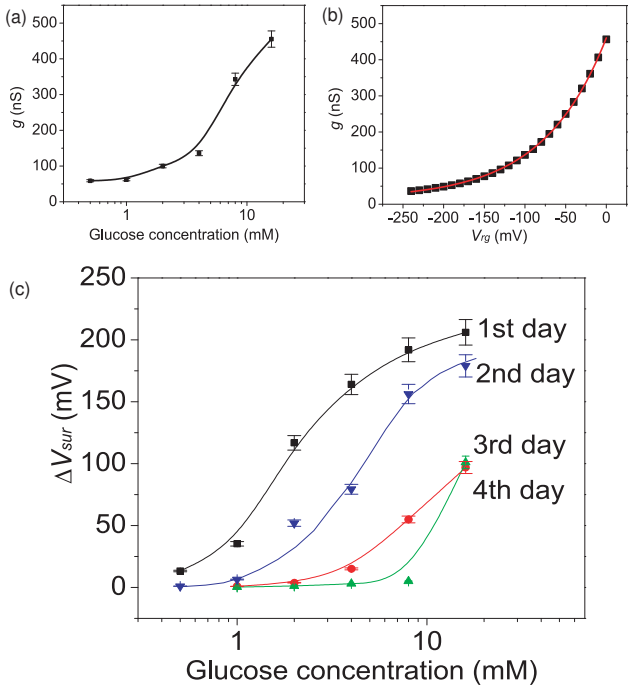}
	\caption{ (a) Device reponse as the function of glucose concentration, taken in the 2nd day. (b) Device response as the function of reference gate voltage in 16 mM glucose solution, taken in the 2nd day.  (c) Calibrated surface potential change vs. glucose concentration, taken at different days. All solution used contents 50 mM of NaCl and 81.8 mM of potassium ferricyanide.}
	\label{fig3}
\end{figure}
To evaluate the device performance and degradation as the function of calibrated surface potential, similar to the macroscopically large ISFET experiments\cite{Bergveld03}, we have measured the calibrated plot of surface potential change. The surface potential $V_{sur}$ is offset from the reference electrode potential because of contact potential drop and electrolytic screening effects. However, the {\it change} in surface potential$V_{sur}$ induced by oxidation reaction is equal to the {\it change} in  reference electrode voltage $V_{rg}$ applied to achieve the same device response in the solution. This is consistent with our previous experiments, in which we showed the equivalence between changing reference electrode voltage and changing analyte concentration in solution\cite{mohanty07}. Fig.~\ref{fig3} shows a plot of the calibrated surface potential change versus glucose concentration. First, we perform real time differential conductance measurements by changing glucose concentration (Fig.~\ref{fig3} (a)). Then, at the same V in the 16 mM glucose solution, we perform a rapid scan of V$_{rg}$ to obtain a similar plot (Fig.~\ref{fig3} (b)). The differential conductance of the device is fitted with an exponential growth function of the reference gate voltage

\begin{equation}
	\left(\frac{dI}{dV}\right)_{V_{rg}}=G_{0}+G_{1}\mbox{ exp}\left(\frac{V_{rg}}{V_\tau}\right),
\end{equation}

\noindent and the value of $V_\tau$ is found to be 77 mV for the listed sample. This fitting parameter implies how the device conductance changes according to surface potential change by adding analyte concentration. Lower value of $V_\tau$ gives higher sensitivity of the device. Using the data from these two measurements, we can plot the calibrated surface potential change versus glucose concentration by converting device response into an equivalent surface potential change. In this step, we convert each differential conductance of the device in different solutions into an equivalent reference gate voltage
by using the curve in Fig.~\ref{fig3} (b), and we count the voltage difference between different solutions. The final plot of calibrated surface potential change as the function of glucose concentration is shown as the second day curve in Fig.~\ref{fig3} (c). From the first day curve, the slope of 160 mV/decade change of glucose concentration is observed. The slope decreases after operation of several days, implying degradation of device performance. This degradation is attributed to a side reaction of the released hydrogen peroxide, which can directly oxidize and deactivate of the enzyme on the device surface. Thus, the application of the device is limited to short-term use. However, long-term use could be achieved by inducing a secondary active electrode near nanowires to eliminate degradation\cite{Rooij94}. 

Taken together, our studies show that the nanowire sensors provide an excellent platform for disposable glucose sensors. In order to implement the sensors for long-term implantable use, a solution to the degradation problem is needed. One suggestion first proposed for macroscale ISFET using a control electrode to counteract the effect of the peroxide appears to be promising. It would be important to extend the performance of nanowire sensors by exploiting a similar idea.

In conclusion, we have fabricated silicon nanowires as glucose biosensor. In the differential conductance measurement, the device response has a linear range for glucose concentration between 0.5 and 8 mM upon initial use. While the calibrated surface potential versus glucose concentration plot
gives the maximum slope of 160 mV/decade. The device also performs with a reasonable response time of 3-5 min The linear glucose detection range, fast response time, and detection limit provide a pathway to fabricate high-density, high-performance nanoscale glucose biosensors that can be integrated with silicon-based circuitry.


\end{document}